# Collision cross section specificity for small molecule identification workflows

Jamie Nunez, Eva Brayfindley, Sean M. Colby, Monee McGrady, Kristin H. Jarman, Ryan S. Renslow[*], Thomas O. Metz[**]

Pacific Northwest National Laboratory, 902 Battelle Blvd, Richland, WA 99354

* Ryan Renslow: ryan.renslow@pnnl.gov

** Thomas O. Metz: thomas.metz@pnnl.gov

ABSTRACT:
The physical-chemical property of molecular collision cross section (CCS) is increasingly used to assist in small molecule identification; however, questions remain regarding the extent of its true utility in contributing to such identifications, especially given its correlation with mass. To investigate the contribution of CCS to uniqueness within a given library, we measured its discriminatory capacity as a function of error in CCS values (from measurement or prediction), CCS variance, parent mass, mass error, and/or reference database size using a multi-directional grid search. While experimental CCS databases exist, they are currently small; thus, we used a CCS prediction tool, DarkChem, to provide theoretical CCS values for use in this study. These predicted CCS values were then modified to mirror experimental variance. By augmenting our search within a library based on mass alone with CCS at a variety of accuracies, we found that, (i) the use of multiple adducts (i.e. alternative ionized forms of the same parent compound) for the same molecule, compared to using a single adduct, greatly improves specificity and (ii) even a single CCS leads to a significant specificity boost when low CCS error (e.g. 1% composite error) can be achieved. Based on these results, we recommend using multiple adducts to build up evidence of presence, as each adduct supplies additional information per dimension. Additionally, the utility of ion mobility spectrometry when coupled with mass spectrometry should still be considered, regardless of whether CCS is considered as an identification metric, due to advantages such as increased peak resolution, sensitivity (e.g. from reducing load on the detector at any given time), improvements in data-independent MS/MS spectra acquisition, and cleaner tandem mass spectral fragmentation patterns.

## INTRODUCTION:

Mass spectrometry (MS) is used across many fields of science for comprehensive analysis of unknown samples due to its high sensitivity and the accuracy of both calculating and measuring monoisotopic molecular masses; however, on its own, accurate mass is not enough to uniquely identify compounds. Often, other instrumentation is used upstream to gather more information (i.e. chemical properties, additional measurements). This can help narrow down potential candidates, thereby increasing the confidence of identifications and decreasing false discovery rates.

Collision cross section (CCS) is a value that can be derived using ion mobility spectrometry (IMS) upstream of a mass spectrometer (IMS-MS). CCS has been shown to be highly reproducible,[2] providing a significant advantage over (or in augmentation to) other measured properties (e.g. retention time and mass fragmentation patterns) that can vary significantly between inter- and even intralab measurements. Studies have shown increased specificity of molecular identifications when using CCS,[3-6] alongside increased sensitivity at the mass spectrometer due to the additional ion separation afforded by IMS.[6]

A study was recently performed by Broeckling et al. [REF] where the CCS of compounds belonging to separate metabolomes (as represented by fooDB and HMDB subsets) were assessed to gain a better understanding of the retention time, mass, and drift time resolutions required to distinguish compounds in such libraries. While this assessment gives great insight to the required resolution of CCS, it is still critical to understand the impact of CCS directly given the available CCS error of a given match, considering multiple adducts (i.e. alternative ionized forms of the same parent compound), and across varying sizes of databases.

In this study, we consider monoisotopic mass and the CCS from zero, one or three adducts. Mass without CCS reflects the importance of mass itself, whether or not multiple adducts are being considered. For example, if two compounds conflict by mass, each of their adducts will as well (assuming they are each able to form such adducts), meaning adducts provide little to no additional structural information, though they do increase the amount of evidence the underlying *formula* is indeed present. Mass with a single CCS (in this paper, we use the protonated adduct) reflects the added specificity provided by CCS per adduct. This is an important case since single ions have their own mass and CCS, making these values easy to consider together during matching and annotation efforts. Mass with three adducts is more complex to achieve in reality, so this represents the best case scenario (when only three adducts are being considered). When it comes to true experimental data, this information would actually come from three separate data points (meaning there would be three sets of mass-CCS pairs coming from different adduct types), making it complicated to truly assign these three data points as belonging to the same parent compound. The complexity of actual identification efforts (e.g. annotation of features, scoring putative identifications, false discovery rates) are beyond the scope of this paper. Here, when three adduct CCS are used, we consider protonated, sodiated, and deprotonated adducts.

Each of these cases are tested for a variety of conditions (CCS error, CCS variance, parent mass, mass error, and/or reference database size). For each of these cases, we assessed the occurrence of conflicts for each compound within each given library and label compounds "unique" if they have zero conflicts.

When comparing these results to any given library and set up, errors here should be compared to the composite error of any two data points being compared. For example, if an analytical measurement has a known CCS error and it is compared against a recorded (i.e. library) CCS value with a known error, these errors would need to be added together to compare to results here (e.g. compare to results in this paper with a 5% error if the two errors associated with these values are 2% and 3%).

## METHODS:

Collision Cross Section (CCS) Prediction. Due to a relatively small number of experimentally-derived CCS (<700 for any given adduct within the Unified CCS Compendium[5] at the time of this study, including protonated, sodiated, and deprotonated adducts) and varying consistency across measurements, we used predictions from the DarkChem prediction tool (v1.0).[7] Because we use only a single source for all CCS values, we take the predicted CCS to be 'ground truth'. We will demonstrate that, even if our DarkChem value variances do not perfectly match true $m/z$-CCS variance, our main conclusions do not change. Using this method, we eliminate issues of consistency across experimental collections or different predictive models. While it does not eliminate error from within the DarkChem model, the bias is assumed to be consistent across all tested databases.

Increased Variance of Predicted CCS. Since this study is reliant on the variance of CCS (which directly correlates with its specificity), it was first ensured that the CCS being used for this analysis has a variance comparable to that of experimental values. CCS tends to correlate closely with mass (Figure S1), especially for predictive models, potentially leading to a lower variance than what is seen experimentally.

Currently, the Unified CCS Compendium[5] (UCCSC) has the largest collection of experimentally measured CCS values available in open-source format. Experimental and predicted (by DarkChem) CCS averages and standard deviations for compounds in this library were compared to create a correction factor for DarkChem values (see Supporting Information, Section S1). For the interested reader, identical analyses were performed using the original (unmodified) DarkChem values (Figure S5, Table S1).

Library Generation. The PubChem database[8] was used to help assess the contribution of CCS in large libraries that include many classes of compounds. CCS values for three adducts (protonated [M+H], sodiated [M+Na], and deprotonated [M-H]) for each compound within the PubChem database were predicted by DarkChem. The library was then downselected to include only masses up to 1,050 Da to focus the results on small molecules (i.e., most primary metabolites, secondary or specialized metabolites, and molecules of interest in environmental / chemical forensics). For additional library exploration, especially to assess the effect of library size and specialized libraries, the ToxCast,[9] Human Metabolome Database (HMDB),[10] Universal Natural Products Database (UNPD),[11] and DSSTox[12] database were also included in a secondary analysis. The ToxCast library is a subset of DSSTox.

Note that for all libraries, we restrict to compounds for which DarkChem is able to make CCS predictions, therefore limiting compounds to those (i) only composed of sulfur, phosphorous, oxygen, nitrogen, carbon, and hydrogen atoms, and (ii) having SMILES of length 100 characters or less.

Matching. For each library, each compound was compared to the rest of the library's compounds and all hits based on mass, given a mass error of 1 ppm or 10 ppm, are initially selected as the mass-discriminated hit list. This list is then reduced by restricting to those compounds within the error window around the query compound's CCS value (or values, in the case of >1 adducts). This provides an estimate of the discriminatory power of CCS at several levels of presumed accuracy. The error windows considered during CCS matching are derived from literature values previously reported for CCS measurements and prediction techniques, where the smallest reported CCS errors are roughly 0.29%,[2] but are usually closer to 3%[6] and are as high as 9%[13]. However, emerging Structures for Lossless Ion Manipulation (SLIM)[14] systems are expected to increase the accuracy and precision of CCS values due to the longer path length traversed during ion mobility measurement. The composite error windows considered here are meant to cover these cases: ±0.1%, ±0.5%, ±1%, and ±5%. We then calculated the percent of unique matches (compounds with no conflicts within the given error windows) and number of conflicts (the number of matches compounds had in their respective library, excluding themselves).

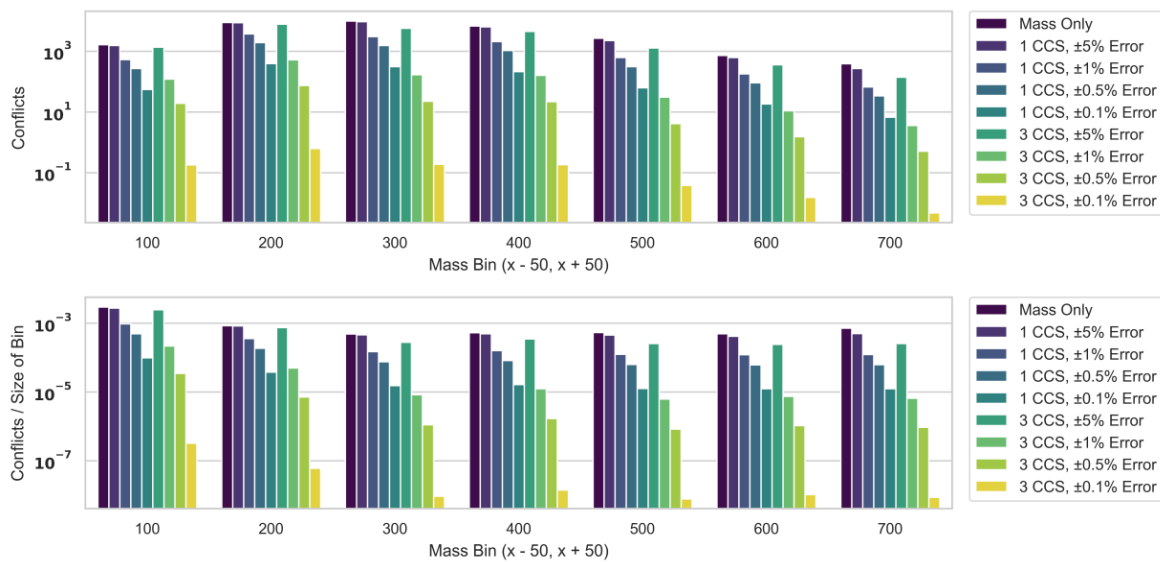

*Figure 1. Conflicts when searching a reference library using parent mass alone or in combination with CCS.* (a) Average number of conflicts. (b) Average number of conflicts divided by the size of the bin (i.e. number of library entries within that 100 Da mass range). One or 3 CCS values were considered, given a ±10 ppm mass error and ±0.1%, ±0.5%, ±1%, or ±5% composite CCS error. The [M+H] adduct is used when searching with a single CCS, and [M+H], [M+Na], and [M-H] adducts are used when searching with three. Results are binned by mass ranges (e.g. if the parent molecule has a mass of 130 Da, then its results are averaged with all data falling in the 50-150 Da range). Colors for the bars were sampled from the viridis colormap. To learn more about the importance of colormap selection, please see Kovesi et al. 2015.[1]

## RESULTS & DISCUSSION:

In **Figure 1**, we present the number of conflicts when searching PubChem using a ±10 ppm mass window (around the parent mass), a ±10 ppm mass window with 1 adduct ([M+H]) CCS, and a ±10 ppm mass window with 3 adduct CCS (protonated [M+H], deprotonated [M-H] and sodiated [M+Na]). With only one adduct and ±5% composite CCS error, CCS does not provide a significant improvement over mass searching alone (3% fewer conflicts on average). With three adducts, the number of conflicts starts to decrease more significantly (25% fewer conflicts on average). When considering 1 vs 3 adduct CCS for the same composite CCS error, a significant advantage is seen with reduced errors. With ±0.1% composite CCS error, we see, on average, 96% and 99.9% fewer conflicts (than ±5% composite CCS error) for one vs three adduct CCS matched, respectively. There is also a balance between the number of available adducts and composite CCS error as well. For example, use of a single CCS with a ±0.1% composite CCS error is comparable to three CCS with a ±1% composite CCS error (average signed percent error between their number of conflicts across mass bins: 24%). After this point, decreasing CCS error does not have as big of an impact on number of conflicts than using 3 adduct CCS. It should be noted again: this finding is considering the identification library on its own. When making identifications, the use of 3 adduct CCS is more complex and can lead to a higher false discovery rate if their relation to the same parent compound is not certain, so this trade-off between accuracy and number of available CCS may not be reflected in practice.

Figure 1a shows CCS specificity is correlated with mass since some bins (e.g. 150-250 Da and 250-350 Da) have a higher number of conflicts on average; however, the number of compounds of these bins is also much higher than in other bins (Figure S6). Figure 1b shows these same results, expect values are divided by the size of their bin. This leads to a more consistent trend across all mass bins except the 50-150 Da mass bin. The variance of CCS assigned across mass bins (calculated from experimentally-available CCS) does not follow this trend (Figure S3). This indicates CCS specificity does not directly correlate with mass, outside the fact that some mass bins are more dense, leading to more conflicts. Upon closer inspection, the higher number of conflicts in the 50-150 Da mass bin appears to be due to compounds not being as evenly distributed in this bin compared to others (Figure S7), meaning a higher density is "felt" by the members falling within this bin.

These findings are reinforced with results for the five open source libraries in **Table 1**, which shows the percent of compounds that are unique (i.e. compounds that have zero conflicts) with varying levels of information and composite errors provided for each match. Only 0.01% were identifiable by mass alone in PubChem. For high accuracy (±0.1% composite error) CCS filtering in addition to mass filtering, 1.5% can be uniquely identified by mass and the [M+H]+ adduct alone, while 77% of compounds can be identified if mass filtering is augmented with high accuracy and three adduct filtering. For the smallest database of these four (ToxCast, containing about 3,000 compounds), 41% of compounds are uniquely identifiable by mass alone, and use of all 3 adducts with high accuracy CCS makes 99.9% of compounds unique.

The same analysis shown in Figure 1 and Table 1 was performed with DarkChem values, before increasing their variance as described in the methods. While the specific resulting values change, the same trends are observed (Figure S5, Table S1). This was also repeated for a mass error of 1 ppm rather than 10 ppm, which again showed the same overall trend (Table S2). In this case, the percent of unique compounds were all within 2%, on average (Table 1 vs. Table S2). For the overall (i.e. not split by mass ranges) average number of conflicts, see Table S3. The 200, 300, and 400 Da mass bins had the most influence on these overall averages due to the large number of compounds falling within these bins (Figure S6).

These results come together to show, compared to mass only filtering, a composite CCS error of 5% does not dramatically increase the percentage of unique compounds (0.5%±0.3% and 1.7%±1.1% more unique compounds for one and three adducts respectively, for the libraries shown in Table 1). However, a larger change is seen in the number of conflicts (5%±2% and 27%±12% less conflicts for one and three adducts respectively, for the libraries shown in Table 1), showing CCS at a 5% error can still help decrease the number of candidates for identification compared to mass-only identification, though additional information would still be required for an unambiguous identification. High accuracy CCS (0.1% composite error) and a match across all 3 CCS further helped with increasing the number of unique compounds compared to mass only identification (85%±15% more unique compounds) as well as decreasing the number of conflicts (96%±1% less conflicts).

Conflicts are further investigated in Figure 2, where a representative example compound, PubChem CID 18061734 (formula: $C_{26}H_{39}N_5O_5$), was selected from the mass 500 range. Here, the parent mass is 501.2951 and the predicted [M+H]+ CCS value is 217.7 $Å^2$. A ±10 ppm window around this mass provides 1201 matches. The standard deviation for [M+H]+ CCS across these matches is 6.9 $Å^2$, with a ±1 standard deviation range in CCS value encompassing 468 of the mass-matched hit list. Also included in Figure 2 are vertical lines indicating where ±0.1%, 1%, and 5% levels are as compared with the standard deviation limits. There are 9 compounds contained within the ±0.1% window, 97 within the 1% window, and 948 within the 5% window.

*Table 1: Percent (%) of compounds that are unique considering varying amounts of evidence. All mass errors are set to ±10 ppm. The number of compounds in each library are provided below each library name.*

| Filter Scheme | Composite CCS Error | ToxCast (~3k) | HMDB (~49k) | UNPD (~143k) | DSSTox (~436k) | PubChem (~51e6) |
|---|---|---|---|---|---|---|
| Mass | NA | 41% | 7% | 3% | 2% | 0% |
| Mass + 1 Adduct ([M+H]) | ±0.1% | 89% | 45% | 36% | 29% | 1% |
|  | ±0.5% | 68% | 22% | 14% | 11% | 1% |
|  | ±1% | 58% | 14% | 9% | 6% | 0% |
|  | ±5% | 42% | 7% | 3% | 2% | 0% |
| Mass + 3 Adducts | ±0.1% | ~100% | 99% | ~100% | 99% | 77% |
|  | ±0.5% | 96% | 65% | 73% | 63% | 8% |
|  | ±1% | 81% | 38% | 37% | 30% | 2% |
|  | ±5% | 45% | 8% | 4% | 3% | 0% |

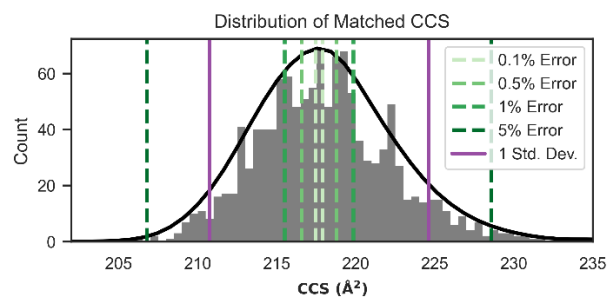

*Figure 2. Gaussian fit to the CCS distribution for ±10 ppm mass-matched set for PubChem CID 18061734.*

## CONCLUSIONS:

Mass spectrometry is a highly sensitive method used to make small molecule identifications but does not provide enough information on its own to enable compound identification due to extensive overlap in mass and even formula space. Additional dimensions (i.e. properties) available alongside mass aids in this gap. One such candidate to improve specificity is collision cross section (CCS), a physical-chemical property that can be consistently measured across setups and labs. In this study, we investigated the number of conflicting compounds within an identification library given varying CCS error, CCS variance, parent mass, mass error, and/or reference database size.

CCS has the potential to provide valuable distinction in addition to mass during compound identification, as long as high accuracy (<1% composite error) is attainable. Currently, most methods (involving experimental and/or predicted values) are working with errors much higher than this (5% composite error or higher). We show here that mass and CCS are not enough to uniquely identify compounds at this level of error, though the use of CCS does provide a reduction in conflicts, thereby aiding in the reduction of false discovery rates. High accuracy CCS increases the percentage of compounds that can be identified with just mass and CCS, and is currently only attainable with measurements on SLIM systems,[14] which are now commercially available (mobilionsystems.com). Increasing the number of adducts also increases the usefulness of CCS, though in reality this comes with challenges due to the increased number of candidate ions being searched for (similar to having a larger library size). When considering any property for more than one adduct, additional adducts must be chosen carefully (e.g. only choosing those that have a high chance of forming given the solution and expected chemical classes therein). Tools such as MSAC[15] help with the selection of adducts to search for and calculation of adduct masses, and there is active research to predict which adduct types should be formed for each type of molecule.[16, 17]

This study also shows the need to assess all measurement (e.g. mass and CCS) errors on every system, and not to assume errors based on out-of-date and/or others' sources. Error is commonly assessed for predicted CCS values as new tools come out, but not for experimental systems, even though errors do appear to vary significantly[2, 6, 18] and may even be higher than previously thought (up to 9%).[13] Error present in a given library (either generated from predictive tools or previously measured) or new measurements (e.g. from an unknown ion being compared to said library) must be understood along with the implications of their magnitudes during identification efforts.

While CCS on their own are limited in their abilities to decrease specificity, IMS is useful beyond CCS generation (e.g. separation, sensitivity, and peak resolution).[3, 6, 19] We also do not account for other properties that can be used for identifications, such as tandem mass fragmentation spectra (MS2), which is frequently used for small molecule identification. The contribution of CCS values to identification when MS2 are available would be a valuable future study. Additional dimensions beyond mass and CCS (such as MS2, relative retention time, cryoIR, and microED) are advised to provide more information and increase identification confidence. A critical target for identification methods is to add as many dimensions as possible, until it is shown to be truly unique for all feasible compounds. Until this is available, completely transparent reporting of dimensions used and their (validated) associated error is critical, along with availability of all data (e.g., identification libraries and raw data) used to make identifications. This will allow re-evaluation in the future as more experimental and predicted data are available, especially for compounds we are not currently looking for in today's studies (i.e., unknown unknowns).

Several drawbacks of this study that can be improved upon in the future are (1) the source of CCS used for analysis, which are currently generated by DarkChem then modified to increase their variance. There may never be enough experimental CCS to perform such a study, but models to generate more accurate CCS are in the works (e.g. DarkChem 2.0, which incorporates 3D structure rather than SMILES input, and ISiCLE 2.0, which will have improved ionization site selection[20] and conformer selection[21]). (2) Using regression to match predicted CCS variance to experimental variance, rather than binned corrections. We do not expect this to change final conclusions since modified CCS predictions used here led to results that were still comparable to unmodified CCS. (3) Evaluating the addition of further properties, such as MS2 and cryoIR.

As IMS gains popularity and more experimental data is available, this analysis should be reviewed to ensure that it aligns with experimental and composite errors as assumed here. The framework developed here can also be adapted to other CCS prediction methods and error windows.

## ASSOCIATED CONTENT

### Supporting Information

The Supporting Information is available free of charge, please contact the primary author for access.

SupportingInformation.doc: additional methods and figures supporting conclusions in this manuscript.

PubChem.zip: Zipped folder containing PubChem database used for analysis (including CCS calculated by DarkChem in Nov 2019).

Other_DBs.zip: Zipped folder containing other database files used for analysis (including CCS calculated by DarkChem in Nov 2019).

## AUTHOR INFORMATION

### Corresponding Author

* Ryan Renslow: ryan.renslow@pnnl.gov
* Thomas O. Metz: thomas.metz@pnnl.gov

### Author Contributions

The manuscript was written through contributions of all authors. All authors have given approval to the final version of the manuscript.

### Funding Sources

National Institutes of Health, National Institute of Environmental Health Sciences

## ACKNOWLEDGMENT

This research was supported by the National Institutes of Health, National Institute of Environmental Health Sciences grant U2CES030170. This work was performed at the Pacific Northwest National Laboratory (PNNL), which is operated for DOE by Battelle Memorial Institute under contract DE-AC05-76RL01830

## ABBREVIATIONS

CCS, Collision cross section; MS, Ion mobility spectrometry; MS, Mass spectrometry; MS2, Tandem mass fragmentation spectra